\documentclass[12pt]{iopart}

\def\lsim{\lesssim}
\newcommand{\om}{\Omega_{\rm M}}
\newcommand{\ola}{\Omega_{\rm\Lambda}}
\usepackage{iopams}
\usepackage{epsfig} 
 
\begin{document}

\title{Constraining photon-axion oscillations using quasar spectra}
\author{Edvard M\"ortsell\footnote{edvard@physto.se} 
        and Ariel Goobar\footnote{ariel@physto.se}, }
\address{Department of Physics, Stockholm University, \\
         S--106 91 Stockholm, Sweden}

\begin{abstract}
Using quasar spectra from the SDSS survey, we constrain the
possibility of photon-axion oscillations as a source of dimming of
high redshift objects. Such a process has been suggested as an
explanation of the apparent faintness of distant Type Ia
supernovae. For most combinations of magnetic field strengths and
plasma densities along the line of sight, large beam attenuations in
broad band filters would also lead to significant differential
attenuation, not observed in the quasar sample.  However, this
conservative study does not exclude the possibility of $\sim 0.1$ mag
dimming of Type Ia supernovae for average plasma densities $n_e \lsim
10^{-8}$ cm$^{-3}$. NIR and MIR spectroscopic studies of high-$z$
sources may be used put further constrains or provide indirect
evidence for the existence of a very light axion.
\end{abstract}


\maketitle

\section{Introduction}
One of the most direct and conceptually simple ways to probe the energy
densities of the universe is through the redshift-distance
relation. The change in the expansion rate caused by the different
energy components can be measured by comparing recession velocities
(i.e., redshifts) at different epochs of the universe (i.e., different
cosmological distances). Type Ia supernovae (SNe) have proved to be
very useful distance indicators since, after proper empirical
corrections, they have small intrinsic dispersion and are luminous
enough to be observed at very large distances. This method has been
used by independent groups to show the existence of a dominant energy
component with negative pressure, e.g., a cosmological constant
\cite{goobar,perlmutter,riess}.
Future supernova (SN) surveys, like the CFHTLS and Supernova/Acceleration Probe
(SNAP) \cite{cfhtls,snapprop} will increase the statistics radically,
especially at large redshifts, thus making the minimization and
control of different systematic effects crucial in order to be able to
take full advantage of the increased statistics.

Concerns have been raised about, e.g., evolution of SN properties,
gravitational lensing, dust absorption and photon-axion oscillations.

Brightness evolution of SNe can be studied by, e.g., comparing SN lightcurves and
spectra at
different redshifts and for different host galaxy types. Studies
indicate that SNe have quite similar properties, even when formed in
very different environments. Therefore cosmic evolution of SN
properties or host galaxy extinction is not believed to affect the conclusion of the universe
being dominated by a negative pressure component \cite{sullivan}.

Systematic effects from gravitational lensing can be diagnosed by
studying the induced asymmetry in the luminosity distribution whereas
intergalactic dust absorption can be disclosed and corrected for from observed
SN colours and spectrophotometry \cite{bias,dustcolor}.

Photons oscillating into very light axions may cause distant SNe to
appear dimmer \cite{csaki,deffayet,csaki2,axion1,fairbairn}. The net
effect could be confused with the existence of a dark energy component
with a negative equation of state. The presence of a significant
component of dark energy has recently been independently inferred from
other cosmological tests, most notable from recent WMAP background
radiation measurements \cite{wmap} in combination with limits on the
Hubble parameter from the HST Key Project
\cite{hstkey} and large scale structure measurements from the 2dFGRS
survey \cite{2dfgrs}.  Therefore the proposed photon-axion mixing does
not replace the need for dark energy, but its equation of state could
be different.

The broad-band luminosity distribution and wavelength dependent
attenuation caused by photon-axion oscillations have not been studied
in any great detail so far. In this paper we examine to what degree it is
possible, despite the complicated wavelength dependence of the effect,
to put limits on the magnitude of the attenuation by studying a sample
of quasar stellar object (QSO) spectra.

\section{Photon-axion oscillations}
Spin 1 particles, as the photon, can mix with zero spin bosons, as the
hypothetical axion, in the presence of a mixing agent that allows for
the conservation of quantum numbers during the process. The mixing
agent could be a magnetic field transverse to the propagation
direction of the photon.

\begin{figure}[t]
  \centerline{\hbox{\epsfig{figure=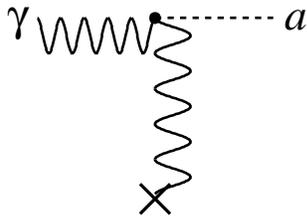,width=0.3\textwidth}}}
  \caption{The conversion between photons and axions in an
  external electromagnetic field.}\label{fig:axion}
\end{figure}

The interaction between the photon and the axion is described by the
Lagrangian
\begin{equation}
	{\cal L}_{\rm int}=\frac{a}{M_{\rm a}}\vec E\cdot\vec B ,
\end{equation}
where $a$ is the axion field, $M_{\rm a}$ is a mass scale determining
the strength of the coupling and $\vec E$ and $\vec B$ are the
electrical and magnetic field, respectively.  The mass scale is given
by
\begin{equation}
	M_{\rm a}\simeq \frac{\pi f_{\rm a}}{\alpha}
\end{equation}
where $f_{\rm a}$ is the decay constant of the axion. Besides the
axion decay $a\rightarrow 2\gamma$, this coupling allows for the
conversion $a\leftrightarrow\gamma$ in an external electromagnetic
field as depicted in Fig.~\ref{fig:axion}.

Due to the interactions with gluons inducing transitions to neutral
pion, axions acquire an effective mass, $m_{\rm a}$, given by
\begin{equation}
\label{eq:mafa}
m_{\rm a}f_{\rm a}\approx m_\pi f_\pi , 
\end{equation}
where $m_\pi =135$ MeV is the pion mass and $f_\pi\approx 93$ MeV its
decay constant. Note however that the axion mass depends on the quark
mass ratios as well as higher-order corrections which are not well
known. In this paper, we assume that there exist an axion with
$m_{\rm a}f_{\rm a}\ll m_\pi f_\pi$.

The most important astrophysical limits on $f_{\rm a}$ are based on
the requirement that the axionic energy loss of, e.g., globular
cluster stars or the core of SN1987A, is not too efficient. A lower
limit on the decay constant is given by $f_{\rm a}\gtrsim 10^9$ GeV
(corresponding to $m_{\rm a}\lsim 10^{-2}$ eV), indicating that
axions, if they exist, are very light and very weakly interacting
\cite{raffeltbook}.

\section{Simulation method}

In order to predict the attenuation due to photon oscillations the
magnetic field strength configuration and the electron plasma along
the line of sight need to be modeled. Earlier work focused on refining
the configuration model. However, the tools used to calculate the
attenuation were too crude to adequately predict the wavelength
dependence of the effect. The most common method used involves a
calculation of the attenuation over a very short distance and then
these small attenuation elements are added to get results over cosmological
distances. Because this method does not to treat the oscillatory
nature of the interactions, it fails to characterize its differential
features along the wavelength axis. A more careful calculation shows
that the effect can vary quite fast with wavelength, especially in the
optical and IR regime.  Thus, the simple approach will introduce
errors larger than that caused by the uncertainties in the magnetic
field strength and electron plasma density. As demonstrated in
Ref.~\cite{axion1}, a full density matrix calculation is necessary
to get reliable results.

\subsection{Magnetic fields and plasma densities}
Only model dependent upper limits on the strength of intergalactic
magnetic fields are available. This has to do with the fact that the
most important observational technique used to trace far-away
intergalactic magnetic fields -- Faraday rotation -- requires
knowledge of the intergalactic electron density, $n_{\rm e}$ as well
as the field configurations
\cite{grasso}. 
The Faraday rotation is the effect of the polarization vector rotating
when light travels through a magnetized medium. The polarization
vector will rotate by an angle \cite{blasi}
\begin{equation}\label{eq:rm}
  \psi =\frac{e^3}{2\pi m_{\rm e}^2c^4}\int_0^{l}
	n_{\rm e}(l)B_{\parallel}(l)\left[\frac{\lambda (l)}{\lambda_{\rm obs}}\right]^2 dl,
\end{equation}
where $\lambda (l)$ is the wavelength at position $l$ along the light
path. Since neither the intergalactic electron density nor the
magnetic field domain size is well known, we have chosen to work with
a simple model where the domain sizes correspond to typical
galaxy-galaxy separations with an average comoving density, $n_{\rm
e}$, with a Gaussian dispersion of 50 \% between domains. In
accordance with Ref.~\cite{fairbairn}, we assume that the magnetic
field is frozen into the plasma and subsequently given by $B_{\rm
0}\propto n_{\rm e}^{2/3}$ with random direction. For each model, we
have tested whether our simulations are consistent with current
Faraday rotation measurements. Since the magnetic domains are small
compared to the total travel length and the magnetic fields have
random directions, the Faraday rotation induced is too small to be
used to rule out any of our investigated models.

\subsection{Density matrix formalism}\label{sec:dmf}
The equation to solve for the evolution of the density matrix $\rho$
is given by \cite{sakurai}
\begin{equation}
  \label{eq:rhoeq}
  {\rm i}\delta_{\rm t}\rho =\frac{1}{2\omega}[M,\rho], 
\end{equation}
with initial conditions
\begin{equation}
  \label{eq:rho0} 
  \rho_{\rm 0}=\left(\begin{array}{ccc}
       \frac{1}{2} & 0 & 0\\
       0 & \frac{1}{2} & 0\\
       0 & 0 & 0
       \end{array}\right).
\end{equation}
Here the three diagonal elements refer to two different polarization
intensities and the axion intensity, respectively and,
\begin{equation}
  \label{eq:M} 
  M=\left(\begin{array}{ccc}
       \Delta_{\perp} & 0 & \Delta_{\rm M}\cos\alpha\\
       0 & \Delta_{\parallel} & \Delta_{\rm M}\sin\alpha\\
       \Delta_{\rm M}\cos\alpha & \Delta_{\rm M}\sin\alpha & \Delta_{\rm m}
       \end{array}\right).
\end{equation}
The quantities appearing in this matrix are given by
\begin{eqnarray}
  \label{eq:terms} 
  \Delta_{\perp} & = & -3.6\times 10^{-25}\left(\frac{\omega}{1\,{\rm eV}}\right)^{-1}
\left(\frac{n_{\rm e}}{10^{-8}\,{\rm cm}^{-3}}\right){\rm cm}^{-1},\nonumber \\ 
\Delta_{\parallel} & = & \Delta_{\perp},\nonumber \\
\Delta_{\rm M} & = & 2\times 10^{-26}\left(\frac{B_{\rm 0,\perp}}{10^{-9}\,{\rm G}}\right)
\left(\frac{M_{\rm a}}{10^{11}\,{\rm GeV}}\right)^{-1}{\rm cm}^{-1},\nonumber \\
\Delta_{\rm m} & = & -2.5\times 10^{-28}
\left(\frac{m_{\rm a}}{10^{-16}\,{\rm eV}}\right)^2
\left(\frac{\omega}{1\,{\rm eV}}\right)^{-1}{\rm cm}^{-1},
\end{eqnarray}
where $B_{\rm 0,\perp}$ is the strength of the magnetic field
perpendicular to the direction of the photon, $M_{\rm a}$ is the
inverse coupling between the photon and the axion, $n_{\rm e}$ is the
electron density, $m_{\rm a}$ is the axion mass and $\omega$ is the
energy of the photon.  The angle $\alpha$ is the angle between a fixed
polarization vector and the perpendicular (projected) magnetic field.

We solve the system of 9 coupled (complex) differential equations
numerically\footnote[1]{We have used the {\tt lsoda} package from Netlib, 
{\tt http://www.netlib.org}}, by following individual light paths through
a large number of cells where the strength of the magnetic field and
the electron density is determined from predefined distributions and
the direction of the magnetic field is random. Through each cell the
background cosmology and the wavelength of the photon are updated, as
are the matrices $\rho$ and $M$.  In all of our simulations, we use a
$[\om =0.3, \ola =0.7, h=0.7]$-cosmology. However, the cosmology
dependence is quite weak.
 
\subsection{Parameter dependence}
In order to study the qualitative behavior of the solutions without
any regard to the polarization state of the photons, we rewrite $M$ as
a $2\times 2$ matrix,
\begin{equation}
  \label{eq:m2D} 
  M^{\rm 2D}=\left(\begin{array}{ccc}
       \Delta & \Delta_{\rm M}\\
       \Delta_{\rm M} & \Delta_{\rm m}
       \end{array}\right),
\end{equation}
where $\Delta =\Delta_{\perp}=\Delta_{\parallel}$ and $\Delta_{\rm M}$
is the component of the magnetic field parallel to the average
polarization vector of the photon beam.  We solve Eq.~(\ref{eq:rhoeq})
for the density matrix $\rho^{\rm 2D}$ with initial conditions
\begin{equation}
  \label{eq:rho02D} 
  \rho_{\rm 0}=\left(\begin{array}{cc}
       1 & 0\\
       0 & 0
       \end{array}\right),
\end{equation}
where the diagonal elements refer to the photon and the axion
intensity respectively.  Assuming a homogeneous magnetic field and
electron density, we can solve the two-dimensional system
analytically. For the $\rho^{\rm 2D}_{11}$ component, referring to the
photon intensity, we get
\begin{eqnarray}
  \label{eq:rho11} 
  \rho^{\rm 2D}_{11}&=&1-\left(\frac{\Delta_{\rm M}}{2\omega\Omega}\right)^2
        (1-\cos{\Omega t}),\nonumber \\
  \Omega &=&\frac{\sqrt{(\Delta -\Delta_{\rm m})^2+2\Delta_{\rm M}^2}}{2\omega}.
\end{eqnarray}
For values close to the typical set of parameter-values as indicated
in Eq.~(\ref{eq:terms}), we can set $\Omega\simeq\Delta /(2\omega )$ to
get
\begin{equation}
  \label{eq:rho11approx} 
  \rho^{\rm 2D}_{11}\simeq 1-\left(\frac{\Delta_{\rm M}}{\Delta}\right)^2
        (1-\cos{\frac{\Delta t}{2\omega}}).
\end{equation}

The oscillation length is of the order $\sim$ Mpc, i.e., comparable to
the size of the domains used in the numerical integration. Note
however that this number is very sensitive to the specific input
parameter values used. In general smaller magnetic domains
(corresponding to smaller cell sizes) yield lower oscillation
probabilities, in accordance with the results of Ref.~\cite{csaki}.

For $m_{\rm a}\gg m_{\rm max}\approx 38\sqrt{n_{\rm e}/(10^{-8}{\rm
cm}^{-3})}\,10^{-16}\,{\rm eV}$, the oscillations are suppressed as
$m_{\rm a}^{-4}$. For $m_{\rm a}\ll m_{\rm max}$, the effect is quite
insensitive to the values of the axion mass which numerical
simulations show is true even for $m_{\rm a}\approx m_{\rm max}$
\cite{axion1}. Thus, in all simulations we set 
$m_{\rm a}=10^{-16}\,{\rm eV}$.

Already from Eq.~(\ref{eq:terms}), it is seen that only the
combination $B_{\rm 0}/M_{\rm a}$ affects the oscillation
probability. Defining $M_{\rm a}^{11}=\frac{M_{\rm a}}{10^{11}\,{\rm
GeV}}$, for small mixing angles, the effect is roughly proportional to
$(B_{\rm 0}/M_{\rm a}^{11})^2$ whereas the effect is rather
insensitive to the exact value of $B_{\rm 0}/M_{\rm a}^{11}$ in cases
of close to maximal mixing. Note that even if the magnetic field
strength $B_{\rm 0}$ would be well constrained from independent
measurements, the relevant quantity $B_{\rm 0}/M_{\rm a}^{11}$ will
not be, due to our lack of knowledge of $M_{\rm a}^{11}$.

Generally the effect is stronger for low values of the electron
density. The value of $n_{\rm e}$ also has a strong impact on the
wavelength dependence of the effect in the respect that higher plasma
densities generally cause a more rapid wavelength dependence. In
Fig.~\ref{fig:oscfig}, we show the median QSO spectrum obtained in
Ref.~\cite{sdsscomp} unaffected by photon-axion oscillations (upper
panel), with [$B_{\rm 0}=10^{-9}, n_{\rm e}\sim 10^{-8}{\rm cm}^{-3}$]
(middle panel) and [$B_{\rm 0}=10^{-9}, n_{\rm e}\sim 10^{-9}{\rm
cm}^{-3}$] (lower panel) for a redshift of $z=1$.  For $B_{\rm
0}=10^{-9}$ G and $n_{\rm e}\sim 10^{-10}{\rm cm}^{-3}$, we have an
oscillation ``wavelength'' of $\sim 1000$ {\AA}. For $n_{\rm e}\sim
10^{-8}{\rm cm}^{-3}$, we have variations over an interval of $\sim 10
- 100$ {\AA} whereas for $n_{\rm e}\gtrsim 10^{-7}{\rm cm}^{-3}$ the
variations occur over intervals of $\sim 1$ {\AA} making them hard to
discern from noise.

\section{Analysis method}
\begin{figure}[t]
  \centerline{\hbox{\epsfig{figure=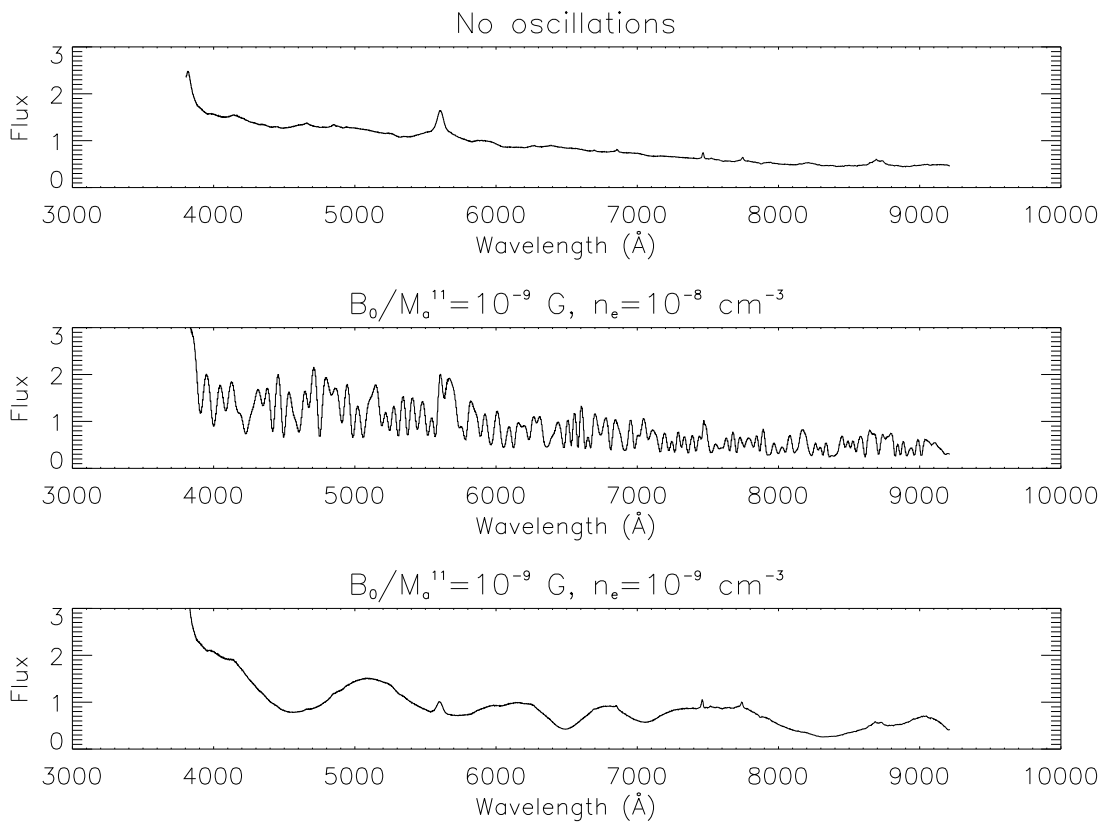,width=\textwidth}}}
  \caption{Median QSO spectrum unaffected by photon-axion oscillations
  (upper panel), with [$B_{\rm 0}=10^{-9}, n_{\rm e}\sim 10^{-8}{\rm
  cm}^{-3}$] (middle panel) and [$B_{\rm 0}=10^{-9}, n_{\rm e}\sim
  10^{-9}{\rm cm}^{-3}$] (lower panel) for a redshift of $z=1$.}
\label{fig:oscfig}
\end{figure}

One of the main difficulties in constraining photon-axion oscillations
is that the effect can look dramatically different for different set
of parameter values, as shown in Fig.~\ref{fig:oscfig}.
Given a set of very similar low-noise, high redshift spectra, we
should nevertheless be able to put useful constraints on this kind of
effect by studying the dispersion around the mean spectrum.

\subsection{The sample}
\begin{figure}[t]
  \centerline{\hbox{\epsfig{figure=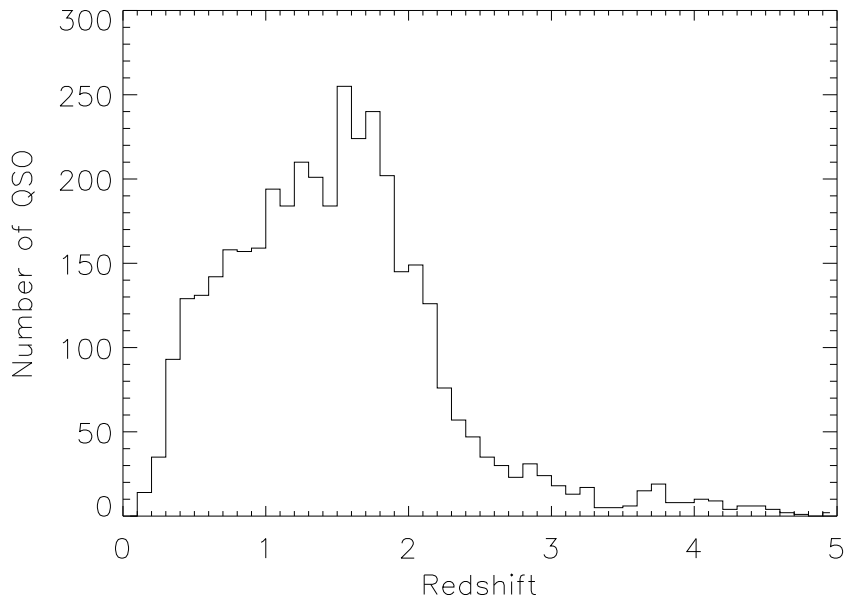,width=0.8\textwidth}}}
  \caption{Redshift distribution of 3814 QSO in the Sloan Digital Sky
  Survey Early Data Release.}  
\label{fig:zdistr}
\end{figure}

\begin{figure}[t]
  \centerline{\hbox{\epsfig{figure=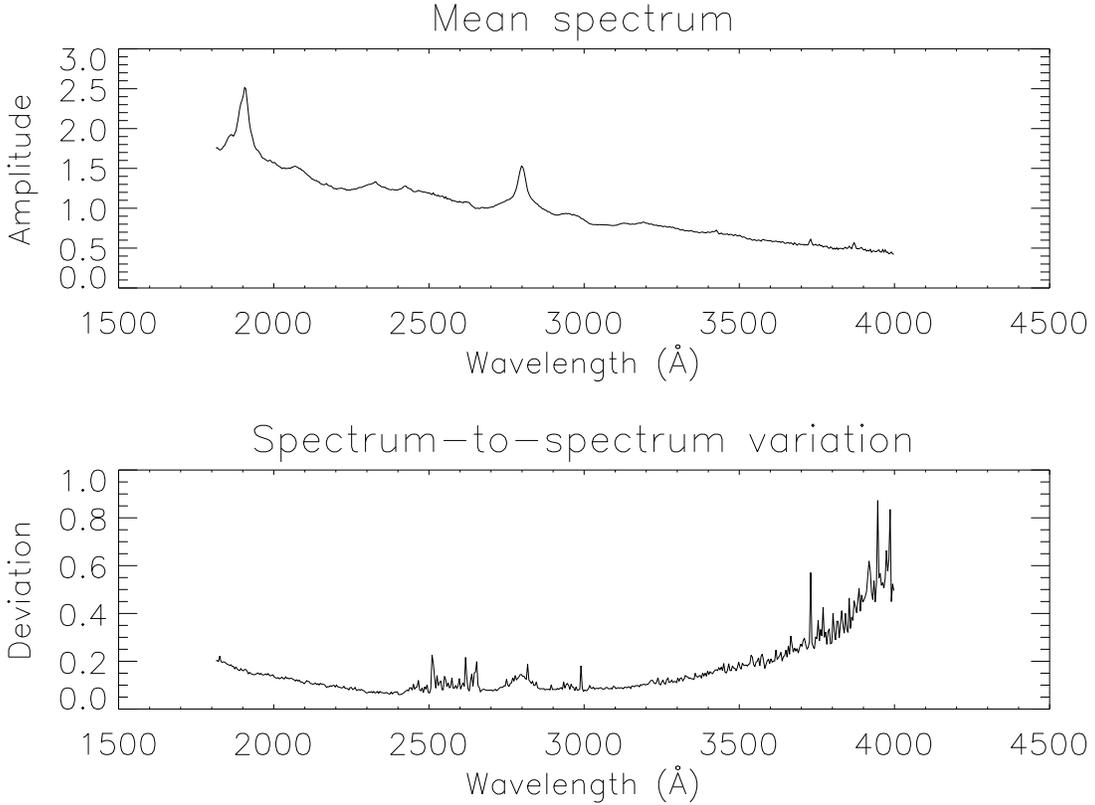,width=\textwidth}}}
  \caption{The mean restframe spectrum (upper panel) and spectrum-to-spectrum
  variation (lower panel) in the redshift bin $1.1<z<1.3$ consisting
  of 383 QSOs with a mean redshift of $z=1.203$.}
\label{fig:sigspec}
\end{figure}

We have used a data sample of 3814 QSO spectra from the Sloan Digital
Sky Survey (SDSS) Early Data Release (EDR) \cite{sdssedr}. The QSOs
have redshifts between $z=0.15$ and $5.03$ distributed according to
Fig.~\ref{fig:zdistr}.
We divide the QSOs into redshift bins of size $\Delta z = 0.2$ in the
interval $0.1<z<2.9$. Each spectrum is smoothed by binning the data into
wavelengths bins of size 4 {\AA} (in the restframe) and
calculating the mean flux in the data points that fall within each
bin. In each redshift bin, we calculate a mean restframe spectrum by
normalizing the flux so that we have a mean flux of one (arbitrary
units) over the common wavelength interval and average over the
smoothed spectra. The spectrum-to-spectrum differential variation depends on the
wavelength and redshift but is lower than 20
\% over large intervals. In Fig.~\ref{fig:sigspec}, we show the mean 
spectrum (upper panel) and spectrum-to-spectrum variation (lower
panel) in the redshift bin $1.1<z<1.3$ consisting of 383 QSOs with a
mean redshift of $z=1.203$.

\subsection{Simulated spectra}
Using the density matrix method described in Sec.~\ref{sec:dmf}, we
have made a large number of Monte Carlo simulations of the effect on
QSO spectra of photons oscillating into axions over cosmological
distances. Our parameters are $B_{\rm 0}/M_{\rm a}^{11}$ and the
plasma density $n_{\rm e}$.  For each set of parameter values and
redshift, we add the effect of photon-axion oscillations to the mean
spectra derived from the real data set (e.g., for $1.1<z<1.3$, upper
panel in Fig.~\ref{fig:sigspec}).

\subsection{Statistical analysis}
In each redshift bin $z$, we calculate the dispersion (in flux),
$\sigma^{r}_{z,i}$, around the mean for each spectrum ($i$). This is
done in two steps, where after the first pass, spectral points more
than 3 sigma off are rejected not to put excess weight on outliers
caused by specific intrinsic features in the spectra. We thus obtain a
{\em distribution} of $\sigma^{r}_{z,i}$ in each redshift bin, i.e.,
one number for each spectrum included in the redshift bin. The routine
is repeated for the simulated data, with a set of parameters
${\vec\theta} = (B_{\rm 0}/M_{\rm a}^{11},n_{\rm e})$ giving a
corresponding set of $\sigma(\vec{\theta})_{z,i}$. Since photon-axion
oscillations only add to the intrinsic dispersion, a very conservative
approach is to rule out any scenario for which $\sigma^{r}_{z,i} <
\sigma(\vec{\theta})_{z,i}$. Comparing the cumulative distributions
$S(\sigma^{r}_{z,i})$ and $S(\sigma(\vec{\theta})_{z,i})$ using the
Kolmogorov-Smirnov (K-S) test and only taking into consideration the
maximum value of the difference $S(\sigma(\vec{\theta})_{z,i}) -
S(\sigma^{r}_{z,i})$ we can use standard statistics (see, e.g.,
\cite{numref}) to obtain a probability $p_z$ that $\sigma^{r}_{z,i}$ 
generally is larger than $\sigma(\vec{\theta})_{z,i}$. By multiplying results
from each redshift bin, we obtain a final probability for the
configuration vector 
\begin{equation}
P(\vec{\theta}) 
= \prod_z{p_z}(\vec{\theta}),
\end{equation} 
that $\sigma^{r} > \sigma(\vec{\theta})$ and can
thus rule out a specific set of input parameter values if, e.g.,
$P(B_{\rm 0}/M_{\rm a}^{11},n_{\rm e})<0.05$.

\begin{figure}[t]
  \centerline{\hbox{\epsfig{figure=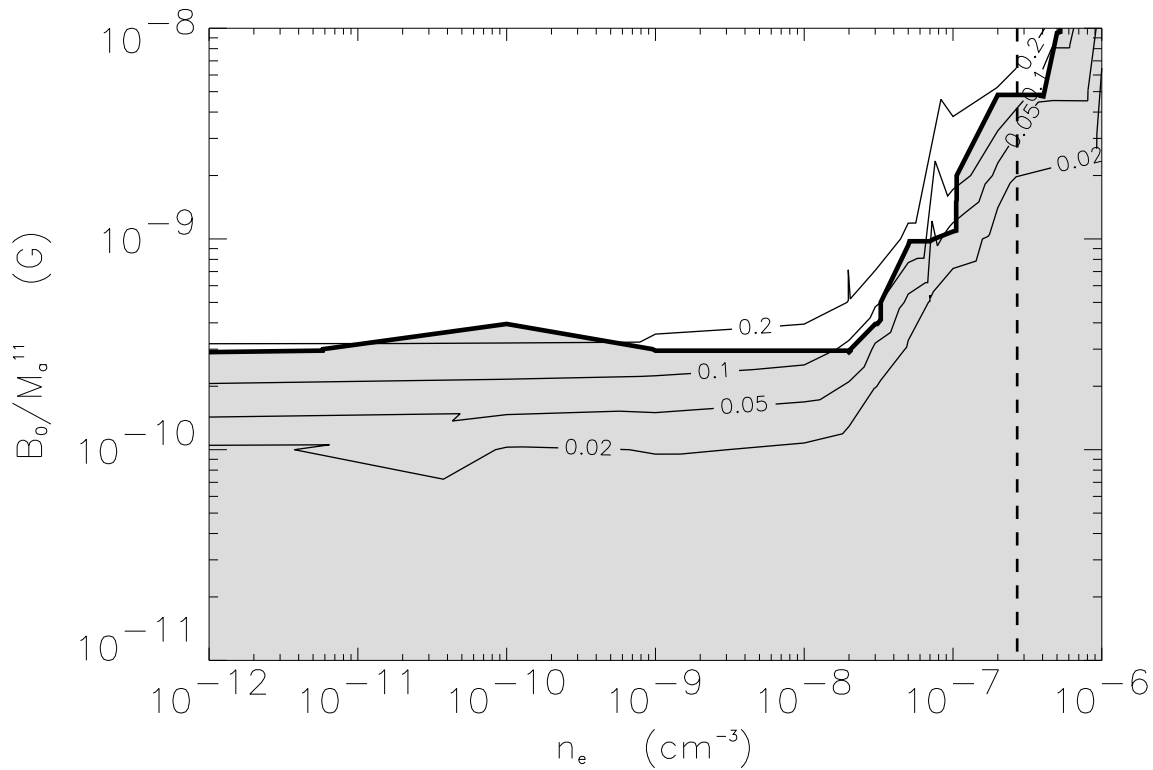,width=\textwidth}}}
  \caption{Combined results from restframe B-band attenuation for Type
  Ia SNe at $z=0.8$ and the K-S test. Contour levels labeled
  [0.2,0.1,0.05,0.02] show the amount of attenuation (in magnitudes)
  while the grey area indicates the 95 \% confidence level allowed
  region from the K-S test. Small scale features are due to low
  resolution in the parameter grid.}
\label{fig:contour}
\end{figure}

We are also interested in what observational effects each set of
parameter values have when integrating over broad band filters. We have
investigated this effect by integrating the attenuation over the
restframe B-band for Type Ia SNe. In Fig.~\ref{fig:contour}, we have
combined the results from this investigation with the results from the
K-S test. The grey area indicates the allowed region in the $[B_{\rm
0}/M_{\rm a}^{11},n_{\rm e}]$-plane from the K-S test using a $P<0.05$
cut, i.e., roughly corresponding to a 95 \% confidence level. Note
however that this is a {\em very} conservative estimation since we have not
taken into consideration the intrinsic variation of QSO spectra. 

The four contour levels labeled [0.2,0.1,0.05,0.02] correspond to the amount
of attenuation caused by photon-axion oscillations when integrating
over the restframe B-band of a set of Type Ia SNe at a redshift of
$z=0.8$. There is a band in the $[B_{\rm 0}/M_{\rm a}^{11},n_{\rm
e}]$-plane where more than 0.05 mag attenuation results beyond what
may be excluded from this analysis. For low values of the plasma
density, the allowed attenuation could be as large as 0.2 mag. Also
indicated (dashed line) is the baryon density inferred from recent
WMAP measurements \cite{wmap}, indicating the maximum allowed value of
the plasma density, $n_{\rm e}\lsim 2.7\cdot 10^{-7}\,{\rm
cm}^{-3}$. To test the effect of the data quality on the results, the
analysis was also done only including spectra with S/N$>$10, with a
somewhat larger exclusion region as a result, emphasizing the
conservative nature of our limits. The small scale features in
Fig.~\ref{fig:contour} are due to low resolution in the parameter
grid.

\begin{figure}[t]
  \centerline{\hbox{\epsfig{figure=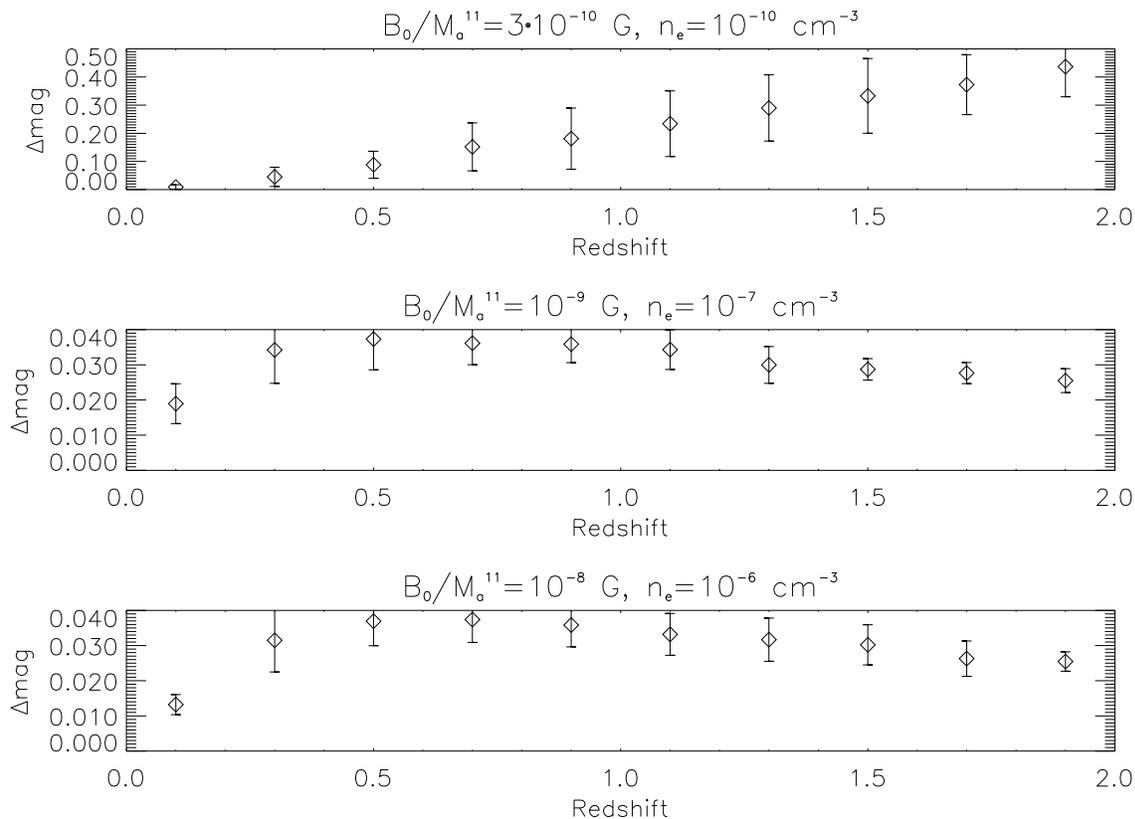,width=\textwidth}}}
  \caption{The B-band photon-axion oscillation attenuation integrated
  over the restframe B-band as function of redshift for [$B_{\rm
  0}/M_{\rm a}^{11}=3\cdot 10^{-10}$ G, $n_{\rm e}=10^{-10}$
  cm$^{-3}$] (upper panel), [$B_{\rm 0}/M_{\rm a}^{11}=10^{-9}$ G,
  $n_{\rm e}=10^{-7}$ cm$^{-3}$] (middle panel) and [$B_{\rm 0}/M_{\rm
  a}^{11}=10^{-8}$ G, $n_{\rm e}=10^{-6}$ cm$^{-3}$] (lower panel).}
\label{fig:axionhubble}
\end{figure}

Previous results indicate that low electron densities are required to
get an attenuation that increases with redshift. However, since the
notion of photon-axion oscillations being solely responsible for the
observed faintness of high redshift SNe seems less and less probable
when independent cosmological tests support the interpretation of a
dominant negative pressure component, we do not restrict our interest
to such scenarios.  Rather, we are trying to constrain the possible
systematic effect for future high precision cosmological tests, e.g.,
the Supernova/Acceleration Probe (SNAP) \cite{snapprop} aiming at a
total error budget of $\Delta{\rm mag} < 0.02$.
In Fig.~\ref{fig:axionhubble} we used the SNOC Monte-Carlo simulation 
package \cite{snoc} to show the amount of attenuation caused by
photon-axion oscillations when integrating over the restframe B-band
of a set of Type Ia SNe as a function of redshift for three sets of
parameter values passing the K-S test. Note that the result for low
plasma densities are very similar to the one for a negative pressure
component and that the overall normalization can be set by
varying the strength of the magnetic fields and/or the photon-axion
coupling strength.

\begin{figure}[t]
  \centerline{\hbox{\epsfig{figure=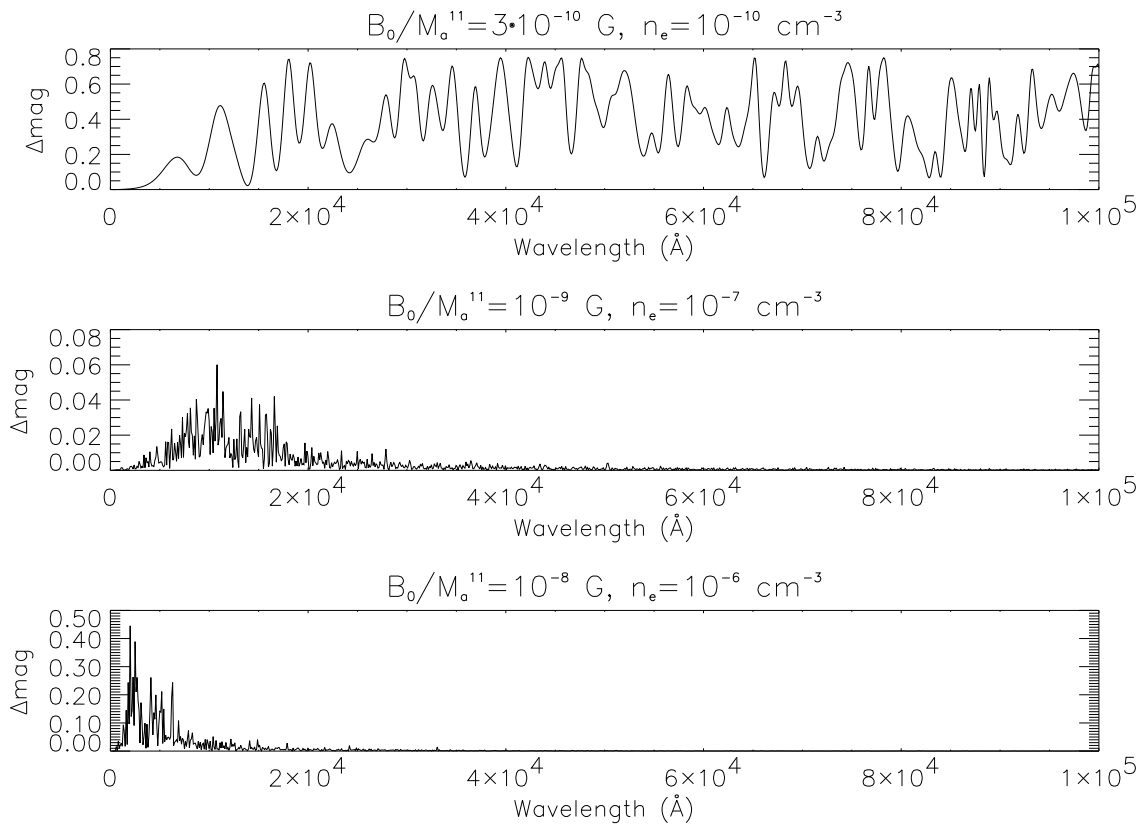,width=\textwidth}}}
  \caption{Photon-axion oscillation attenuation over a wide energy
  range for [$B_{\rm 0}/M_{\rm a}^{11}=3\cdot 10^{-10}$ G, $n_{\rm
  e}=10^{-10}$ cm$^{-3}$] (upper panel), [$B_{\rm 0}/M_{\rm
  a}^{11}=10^{-9}$ G, $n_{\rm e}=10^{-7}$ cm$^{-3}$] (middle panel)
  and [$B_{\rm 0}/M_{\rm a}^{11}=10^{-8}$ G, $n_{\rm e}=10^{-6}$
  cm$^{-3}$] (lower panel).}
\label{fig:checkwide}
\end{figure}

A possible concern is what implications photon-axion oscillations can
have for the CMB observations. As noted by previous authors, assuming
that the magnetic fields form at relatively low redshifts, the
background radiation will be redshifted to such long wavelengths that
the effect becomes negligible. This can be seen by studying
Eq.~(\ref{eq:rho11}) for small photon energies. In doing this, we can
also see that for infrared (IR) energies, we do have a sizable effect
for low plasma densities which can be used to put constraints complementary
to what we have derived using only optical spectra, see Fig.~\ref{fig:checkwide}.  

\subsection{Polarization}
\begin{figure}[t]
  \centerline{\hbox{\epsfig{figure=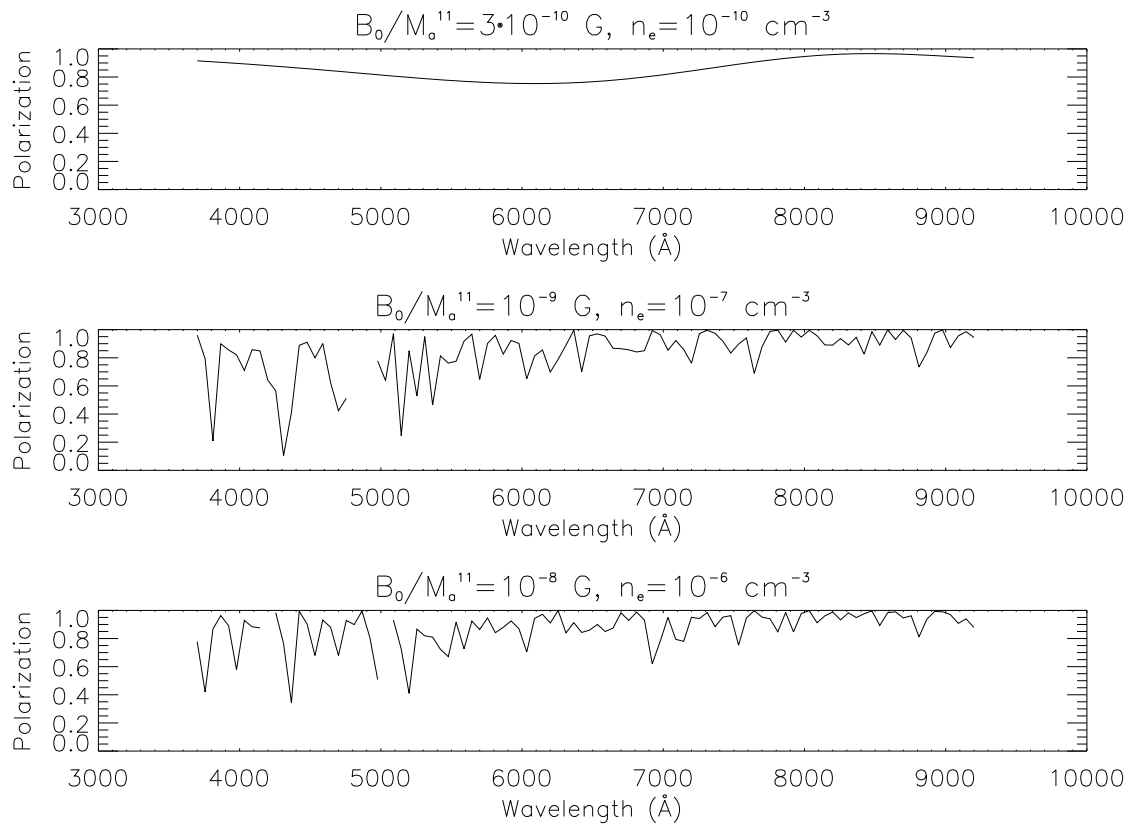,width=\textwidth}}}
  \caption{The observed polarization as a function of wavelength for
  an intrinsically fully polarized source at a redshift of $z=1$.}
\label{fig:polfig}
\end{figure}

Due to the mixing between the different polarization states and the
axion, polarized light might lose some of the polarization when
travelling cosmological distances. In contrast to Faraday rotation,
this effect can be large even when having a large number of magnetic
domains with uncorrelated field strength and direction. Since the
amount of depolarization depends on the values of $B_{\rm 0}/M_{\rm
a}^{11}$ and $n_{\rm e}$, observations of high redshift polarized
sources can be used to put limits on these parameter values. However,
the highest degree of polarization observed for high redshift sources
is $\sim 20$ \% (see, e.g., Ref.~\cite{fairbairn} for a short
compilation of observations) and by comparing this values with the
results depicted in Fig.~\ref{fig:polfig} it is clear that these
observations can not be used to add any additional constraints on the
magnetic field strength and/or plasma density.

\section{Discussion}

The lack of oscillatory dependence in the differential attenuation of
high-$z$ QSO optical light allows us to set severe limits on the
parameter space for photon-axion oscillations. However, the present
very conservative analysis does not exclude that such effects may be
affecting the measured flux of $z \sim 1$ objects at the $\sim$10 \%
level. From Fig.~\ref{fig:axionhubble}, it is clear that assuming a
low electron plasma density ($n_{\rm e}\sim 10^{-10}\,{\rm cm}^{-3}$),
a low mass axion with sufficient coupling strength can imitate the
effect of a negative pressure component with respect to the observed
luminosity of high redshift sources. Since there are independent
cosmological tests supporting the notion of a dominant negative
pressure component in the Universe today, such a scenario seems
unlikely. However, to ensure the use of, e.g., high-$z$ Type Ia
supernovae for precise estimates of cosmological parameters, this
issue needs to be addressed further. This could be done by extending
our present study of quasar spectra to also include IR information.
In the future, instruments like JWST or SNAP may be able to rule out
this scenario or provide independent evidence for the existence of a
very light axion.

\section*{Acknowledgments}
The authors would like to thank Georg Raffelt for helpful comments. 
A.G. is a Royal Swedish Academy Research Fellow supported by a grant
from the Knut and Alice Wallenberg Foundation.

\end{document}